\documentclass[useAMS,usenatbib]{mn2e}

\usepackage{graphicx}
\usepackage{epstopdf}
\usepackage{natbib}
\usepackage{xcolor}

\usepackage{amsmath}

\def \figwidth {147mm}

\title[The inner cavity of the circumnuclear disc]{The inner cavity of the circumnuclear disc}
\author[M. Blank, M. R. Morris, A. Frank, J. J. Carroll-Nellenback and W. J. Duschl]{M. Blank$^{1}$\thanks{E-mail:
mblank@astrophysik.uni-kiel.de (MB); \newline morris@astro.ucla.edu (MRM)}, M. R. Morris$^{2}$\footnotemark[1],
A. Frank$^{3}$, J. J. Carroll-Nellenback$^{3}$ \newauthor and W. J. Duschl$^{1,4}$\\
$^{1}$Institut f\"{u}r Theoretische Physik und Astrophysik,
      Christian-Albrechts-Universit\"{a}t zu Kiel, Leibnizstr. 15, 24118 Kiel, Germany \\
$^{2}$Department of Physics and Astronomy, University of California, Los Angeles, CA 90095-1547, USA \\
$^{3}$Department of Physics and Astronomy, University of Rochester, Rochester, NY 14627, USA \\
$^{4}$Steward Observatory, The University of Arizona, 933 N. Cherry Ave., Tucson, AZ 85721, USA}

\begin{document}

\date{\today}

\pagerange{\pageref{firstpage}--\pageref{lastpage}} \pubyear{2016}

\maketitle

\label{firstpage}

\begin{abstract}
The circumnuclear disc (CND) orbiting the Galaxy's central black hole is a reservoir 
of material that can ultimately provide energy through accretion, or form stars in the presence
of the black hole, as evidenced by the stellar cluster that is presently located at the CND's centre.  
In this paper, we report the results of a computational study of the dynamics of the CND.
The results lead us to question two paradigms that are prevalent in previous research on the Galactic Centre.
The first is that the disc's inner cavity is maintained by the interaction
of the central stellar cluster's strong winds with the disc's inner rim, and second, that
the presence of unstable clumps in the disc implies that the CND is a transient feature.
Our simulations show that, in the absence of a magnetic field, the interaction of the wind with the inner disc rim
actually leads to a filling of the inner cavity within a few orbital time-scales, contrary to previous expectations.
However, including the effects of magnetic fields stabilizes the inner disc rim against rapid inward 
migration. Furthermore, this interaction causes instabilities that continuously 
create clumps that are individually unstable against tidal shearing.
Thus the occurrence of such unstable clumps does not necessarily mean that the disc is itself a transient phenomenon.
The next steps in this investigation are to explore the effect of the magnetorotational
instability on the disc evolution and to test whether the results presented here persist
for longer time-scales than those considered here.
\end{abstract}


\section{Introduction}
It is now widely accepted that the Galactic Centre contains a central black hole having a mass of about
$4\cdot10^{6}\,\text{M}_{\odot}$ \citep{2008_Ghez_Salim_Weinberg, 2009_Gillessen_Eisenhauer_Trippe, 2009_Schoedel_Merritt_Eckart, 2014_Yelda_Ghez_Lu}, 
and  is surrounded by a gaseous disc, the so-called circumnuclear disc (CND).
The disc has an inner cavity with a radius of $\sim$1.4\,pc, an outer radius of 4-7\,pc, 
densities of $\sim10^{5}\,\text{cm}^{-3}$, and temperatures of a few 100\,K
\citep{1982_Becklin_Gatley_Werner, 1985_Genzel_Crawford_Townes, 1987_Guesten_Genzel_Wright,
1990_Sutton_Danchi_Jaminet, 1995_Zylka_Mezger_Ward-Thompson, 1999_Morris_Ghez_Becklin}.

The CND is observed to consist of a clumpy medium, where the clumps have sizes of 0.1-0.2\,pc
and densities of $10^{5}$-$10^{6}\,\text{cm}^{-3}$ \citep{1989_Genzel, 1989_Mezger_Zylka_Salter, 1993_Marr_Wright_Backer,
2012_Martin_Martin-Pintado_Montero-Castano, 2013_Lau_Herter_Morris}.
The critical Roche density at 2\,pc is $\sim10^{7}\,\text{cm}^{-3}$ \citep{1987_Guesten_Genzel_Wright},
thus these clumps are not stable against tidal shearing and get destroyed within a dynamical time.
This short lifetime of the clumps has previously led some authors to the conclusion that the CND
cannot be a permanent feature and thus must be an ephemeral phenomenon
\citep{2012_Martin_Martin-Pintado_Montero-Castano, 2012_Requena-Torres_Guesten_Weiss}.
However, other investigations suggest clump densities of about 3-4$\cdot\,10^{7}\,\text{cm}^{-3}$ and thus argue
for a non-transient phenomenon with a lifetime of $\sim10^7\,\text{yr}$ \citep{2005_Christopher_Scoville_Stolovy,
2009_Montero-Castano_Herrnstein_Ho}.
\citet{2001a_Vollmer_Duschl, 2001b_Vollmer_Duschl} also argue for a non-transient phenomenon.
They developed an analytical model for a clumped disc; their clumps are stabilized by rotation
and magnetic fields and thus are stable against tidal disruption for radii larger than 2 pc.

The mass of the CND has been variously estimated to be within a factor of a few of $\sim10^{4}\,\text{M}_{\odot}$,
if one does not assume that the clumps are virialized
\citep{1985_Genzel_Crawford_Townes, 1986_Serabyn_Guesten_Walmsley, 1989_Mezger_Zylka_Salter}.
The magnetic field in the CND is also relatively strong; it has field strengths of the order of a milligauss and a toroidal orientation
\citep{1988_Werner_Davidson_Morris, 1993_Hildebrand_Davidson_Dotson, 1995_Plante_Lo_Crutcher, 2005_Bradford_Stacey_Nikola}.
\citet{1990_Wardle_Konigl} developed a model for the CND's magnetic field, where the toroidal
field is created by differential rotation from a large-scale axial magnetic field.

The CND's inner cavity hosts a central stellar cluster that collectively launches an outflow
with velocities of about $700\,\text{km}\,\text{s}^{-1}$ and an overall mass outflow rate of
about $5\cdot10^{-3}$\,M$_{\odot}\,\text{yr}^{-1}$.
The outflow shows roughly spherically symmetric radial motions and is
thus not dominated by a rotational component \citep{1987_Geballe_Wade_Krisciunas}.
It is generally assumed that the CND's inner cavity is maintained by the interaction
of this outflow with the inner rim of the CND \citep[e.g.,][]{1987_Guesten_Genzel_Wright, 2010_Genzel_Eisenhauer_Gillessen}.
However, the inner cavity is not completely void of gas, but contains some streams of ionized gas,
collectively known as Sgr A West, or the \textquotedblleft minispiral\textquotedblright\,
\citep{1983_Ekers_van-Gorkum_Schwarz, 1983_Lo_Claussen, 1991_Lacy_Achtermann_Serabyn, 2003_Scoville_Stolovy_Rieke, 2004_Paumard_Maillard_Morris}.
In an early paper on the central cavity \citet{1989_Duschl} argued for the inner rim of the CND being located where
the gravitational potential of the extended mass distribution of the galaxy becomes more
important in comparison to the (almost) pointlike potential of the central black hole.

In this paper we present magnetohydrodynamical (MHD) simulations of the CND to
investigate the interaction between the central outflow and the inner rim of the CND.
Our model includes the gravitational field of the central black hole, a disc with an inner cavity,
an initially toroidal magnetic field and the outflow from the central stellar cluster, assumed to be spherical.

In Section~\ref{sec:comp} we give an overview of the computational methods we are using and 
of the initial conditions and parameters of our simulations.
In the following sections we investigate the interaction of the inner rim and the outflow, with and without
the effects of magnetic fields. We furthermore compare our results with simulations that do not include an outflow,
and give a quantitative estimate of the time-scale of the collapse of the inner cavity and the resulting mass
accretion rate due to this collapse. In the last section we conclude with a summary of our findings.
In Appendix~\ref{sec:numvisc} we furthermore estimate the numerical viscosity in our simulations and
show that our results are not substantially affected by this effect.

\section{Computational methods}\label{sec:comp}
We use the adaptive-mesh-refinement MHD code AstroBEAR \citep{2009_Cunningham_Frank, 2013_Carroll-Nellenback_Shroyer_Frank},
which is a high-resolution shock-capturing Eulerian grid code.
It solves the equations of hydrodynamics and the equations of ideal magnetohydrodynamics using the constrained
transport approach for integrating the induction equations to maintain the solenoidality of the magnetic field.
Because the CND is almost entirely molecular, consisting predominantly of H$_2$,
we use the equation of state of an ideal diatomic gas with $\gamma = 1.4$.

As the mass of the central black hole exceeds the mass of the disc by $\sim$ two orders of magnitude
we do not consider the gravitational force of the disc.
We furthermore do not have any physical viscosity implemented.
Our simulations span no more than 13 orbital time-scales, which is very short compared to the viscous time-scale
\citep[which is about 100-1000 orbital time-scales,][]{2000_Duschl_Strittmatter_Biermann}.
Thus we do not expect viscous processes to play a crucial role in our scenario.

\subsection*{Cooling}
As molecular hydrogen is the most abundant molecule in the CND \citep{1985_Harris_Jaffe_Silber},
we assume that the disc consists entirely of H$_2$.
We use the cooling function developed by \cite{1983_Lepp_Shull}, who calculated radiative cooling of H$_2$
using four vibrational lines, each with 21 rotational states.
The cooling rate was fitted by an analytical function that gives the cooling rate as a function of the temperature
and density of the gas.
Using this approach we obtain disc temperatures of a few hundred Kelvin that are in accordance with observations.

\subsection*{Outflow}
To model the outflow from the central stellar cluster we define a spherical region with a radius of $0.5\,\text{pc}$, and
in every timestep of the code we set the hydrodynamic variables inside this region to specific values, namely to a density
of $100\,\text{cm}^{-3}$, a temperature of $10^6\,\text{K}$ and a velocity of $700\,\text{km}\,\text{s}^{-1}$ pointing radially outwards.
These parameters give a global outflow rate of $5.5\cdot10^{-3}$\,M$_{\odot}\,\text{yr}^{-1}$ through the outer surface 
of this region.

\subsection*{Initial conditions and parameters}\label{sec:ic}
The computational domain has the dimensions $20{\times}20{\times}10$\,pc$^3$ and consists initially of
$32{\times}32{\times}16$ cells, corresponding to a resolution of $0.625\,\text{pc}$.
We allow up to 5 levels of refinement, corresponding to a maximum resolution of about $0.02\,\text{pc}$.
The central black hole is assumed to have a mass of $4.2 \cdot 10^6\,\text{M}_{\odot}$. We do not account for accretion,
i.e., for mass growth of the black hole, because the numerical setup of the outflow from the central stellar cluster
prevents any kind of mass inflow towards the black hole
and the time-scales of interest are short enough to allow for this approximation.

\begin{figure}
 \makebox[0.47\textwidth][c]{\includegraphics[width=73mm]{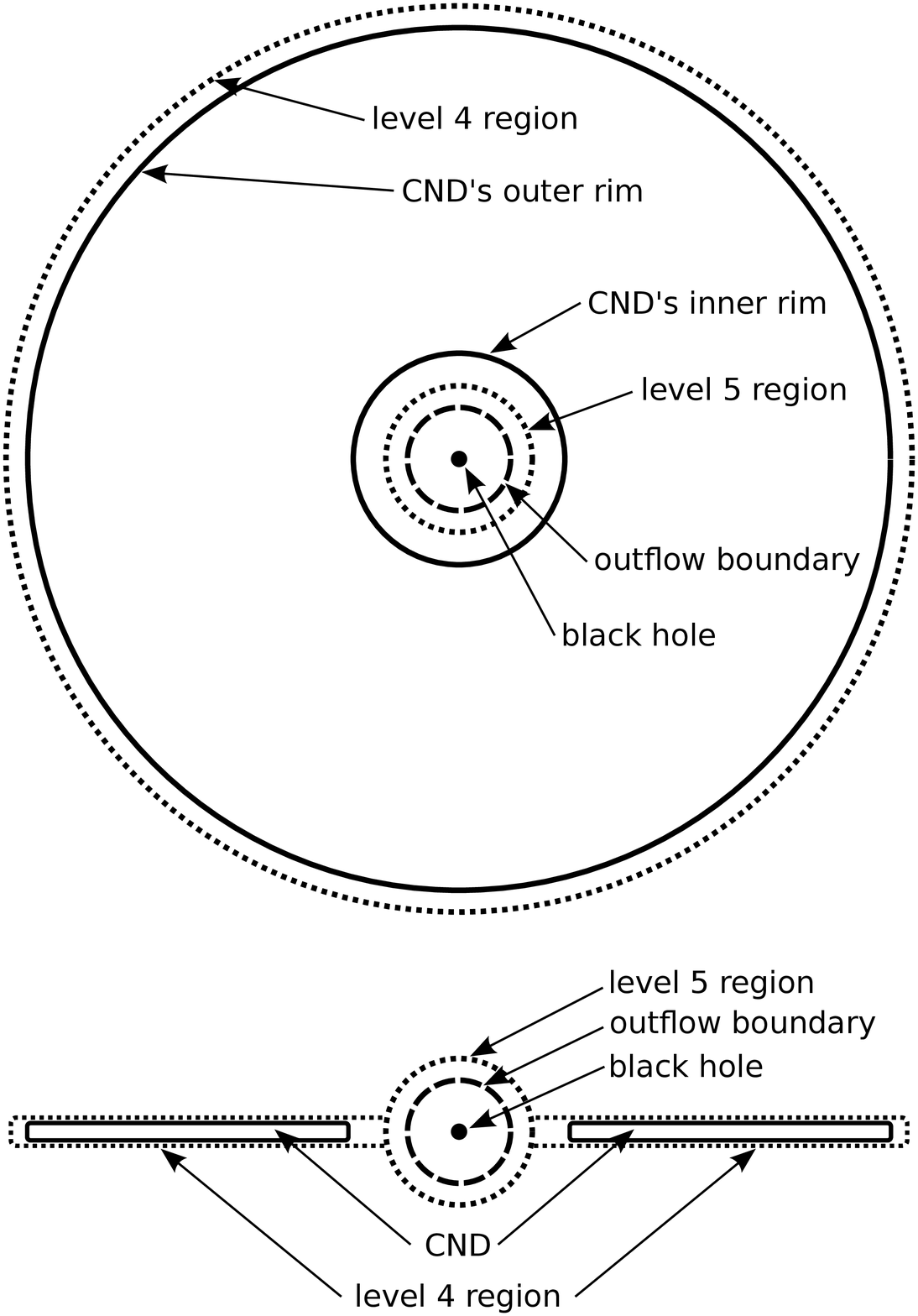}}
 \caption{Initial setup of our simulations. Top: face-on view; bottom: edge-on view.}
 \label{fig:setup}
\end{figure}

The disc  initially has an outer radius of $4\,\text{pc}$, an inner cavity with a radius of $1\,\text{pc}$,
a thickness of $0.2\,\text{pc}$ and a uniform density of $2\cdot10^{5}\,\text{cm}^{-3}$,
which gives a mass of $4.5\cdot10^{4}\,\text{M}_{\odot}$.
The initial temperature is $300\,\text{K}$ throughout the disc.
The disc is embedded in an ambient medium with a density of $1\,\text{cm}^{-3}$ and a temperature of $10^6\,\text{K}$.
The initial magnetic field is constrained to be present only within the disc, and to have a toroidal
geometry and a uniform field strength of $1\,\text{mG}$.
At the location of the disc we always demand a resolution of at least level 4 ($0.04\,\text{pc}$).
To properly resolve the outflow we use the maximum resolution (level 5) within a radius of $0.7\,\text{pc}$
around the black hole.
In Fig.~\ref{fig:setup} we show the initial setup of our simulations.

\section{Interaction of the outflow with the CND's inner rim}\label{sec:interact}
\begin{figure*}
 \makebox[\textwidth][c]{\includegraphics[width=\figwidth]{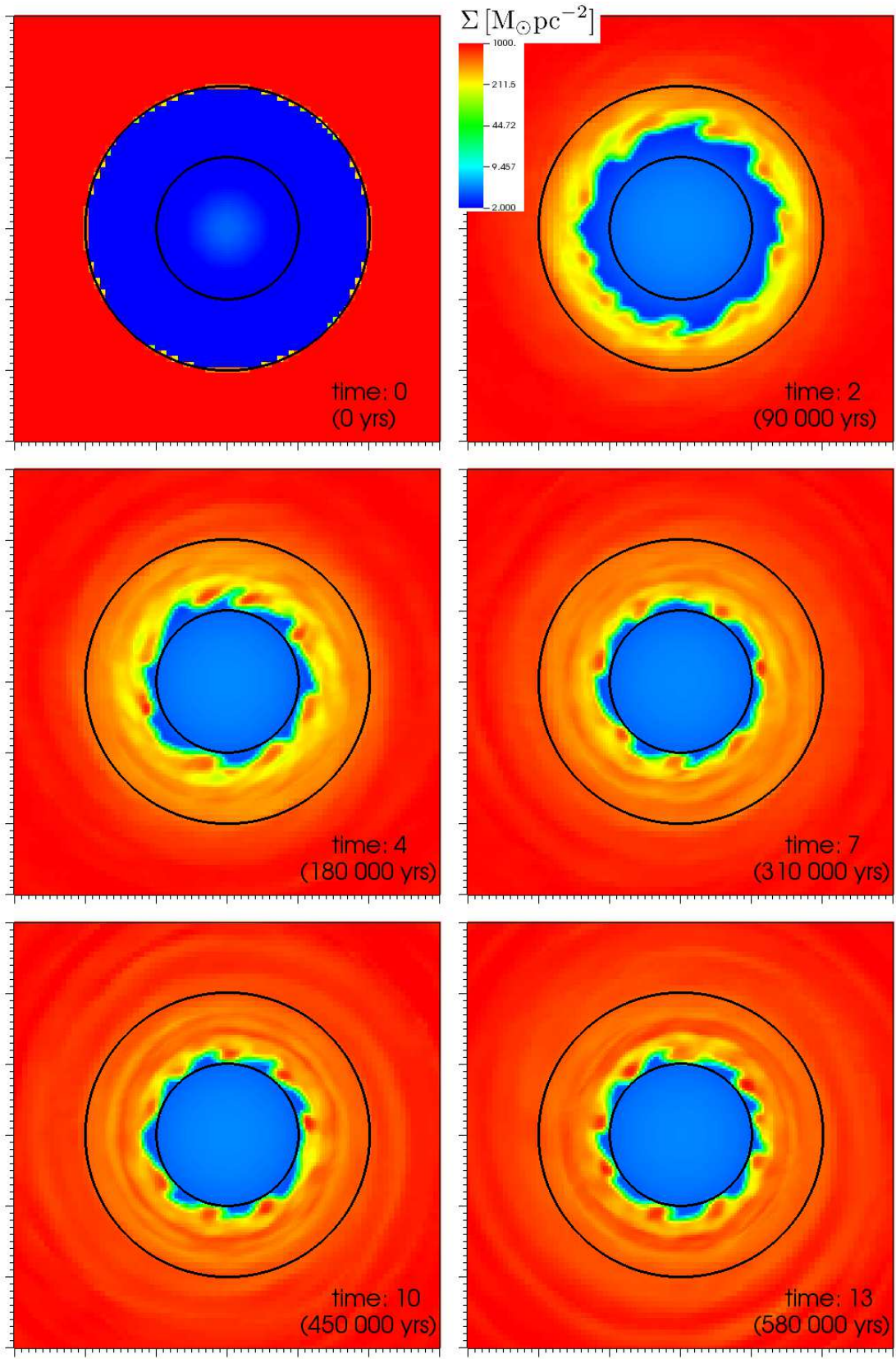}}
 \caption{Surface density of the inner disc for the simulation without magnetic field at different times.
          Each snapshot shows a region of size $3{\times}3$\,pc.
          The time is given in units of the orbital time-scale at 1\,pc, which is $4.5\cdot10^4\,\text{yr}$.
          The outer black circle marks the location of the disc's initial inner rim,
          the inner black circle marks the outer boundary of the region from which the outflow is launched.
          Note that the color scale is logarithmic.}
 \label{fig:sigma_woB}
\end{figure*}

The first simulation does not include any magnetic field. The results are presented in 
Fig.~\ref{fig:sigma_woB}, showing the surface density of the disc's inner region at different times.
As no physical viscosity is implemented in our simulations one might expect that the disc's inner rim 
is pushed outwards by the outflow, or at least that the inner cavity maintains a stable configuration
if the outflow's momentum is not strong enough to push the inner rim outwards.
But Fig.~\ref{fig:sigma_woB} shows that within four orbital time-scales the density
in the inner cavity increases by more than two orders of magnitude, from about $1\,\text{M}_{\odot}\text{pc}^{-2}$
to 400-600 $\text{M}_{\odot}\text{pc}^{-2}$.
Only very close to the outflow boundary, within a distance of 0.1\,pc, 
there are still some low-density regions in the form of a \textquotedblleft cogwheel pattern\textquotedblright.

Thus the inner rim quickly moves inwards until it has reached the location of the outflow boundary.
We emphasize that due to the numerical setup no material enters the inner 0.5\,pc where the outflow
is launched, thus the inner rim stops migrating inwards at this radius.
Contrary to what one would expect and what has been previously assumed in the literature,
the interaction of the outflow and the disc's inner rim causes the latter to shrink.
The outflow from the central stellar cluster does not maintain the inner cavity, it paradoxically causes it to fill in.

The extraction of angular momentum in the absence of physical viscosity can be explained as follows:
by interacting with the disc's inner rim the outflow is adding mass to the inner rim.
Thus its mass increases, but its angular momentum does not change, as the outflow does not carry any angular momentum.
As a result, the specific angular momentum, the angular momentum per unit mass, decreases.
Thus angular momentum can be efficiently extracted from the disc by transferring it to the outflowing wind.
In the next section we give a quantitative estimate for the time-scale of the angular momentum loss
and the resulting mass accretion rate.

The idea that outflows can principally lead to accretion of gas has been discussed by \citet{2013_Dehnen_King}.
In their scenario an AGN-driven wind increases the gravitational energy of the surrounding gas and thus
decreases its periapsis, and second, gas parcels accelerated by an outflow to different eccentric orbits
will collide upon their return to periapsis, and thereby lose angular momentum.

The central stellar cluster's outflow can potentially have additional effects on the CND:
\citet{2013_Zubovas_Nayakshin_King} argue that AGN outflows are capable of triggering starbursts
in the gas discs of their host galaxies, and \citet{2012_Zubovas_Nayakshin} argue for AGN outlows
being capable of removing the entire disc, or at least relocating it to larger radii.
Although the central stellar cluster's outflow is presently much weaker than the AGN outflows discussed in these papers, 
these effects could become important in a later phase of the CND's activity cycle \citep{1999_Morris_Ghez_Becklin}.

Furthermore, the interaction of the outflow with the accretion disc's inner rim creates instabilities at their interface.
The nature of this instability is probably a combination of Kelvin-Helmholtz and Richtmyer-Meshkov instabilities:
our model consists of two fluids (the outflow and the disc) with different densities separated by
an interface and with a velocity difference across the interface.
The only difference with the classical Kelvin-Helmholtz instability is that the velocity of one of
the fluids (in this case the outflow, i.e. the lighter fluid) is supersonic and perpendicular to the interface.
This in turn implies the occurrence of a shock wave and thus suggests the onset of a Richtmyer-Meshkov instability.
We will further investigate this instability and the mechanism that causes the extraction of angular momentum,
and the interplay between them, in our future work.

\begin{figure*}
 \makebox[\textwidth][c]{\includegraphics[width=\figwidth]{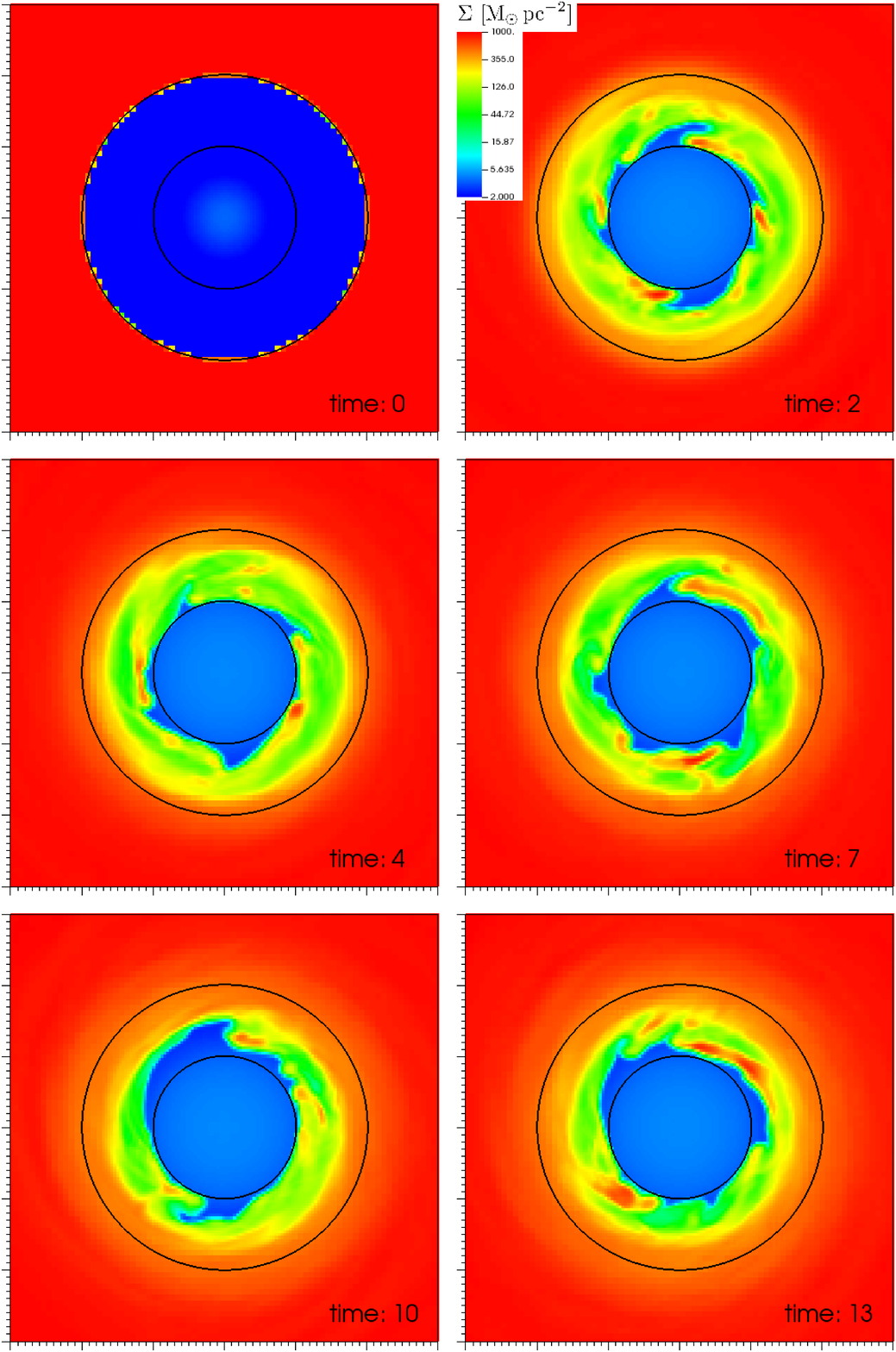}}
 \caption{Surface density of the inner disc for the simulation that includes magnetic fields at different times.
          Each snapshot shows a region of size $3{\times}3$\,pc.
          The time is given in units of the orbital time-scale at 1\,pc, which is $4.5\cdot10^4\,\text{yr}$.
          The outer black circle marks the location of the disc's initial inner rim,
          the inner black circle marks outer boundary of the region from which the outflow is launched.
          Note that the color scale is logarithmic.}
 \label{fig:sigma}
\end{figure*}

Our second simulation includes magnetic fields, as described in Section \ref{sec:ic}.
Fig.~\ref{fig:sigma} shows again the surface density of the disc's inner region at different times.
Clumps and streams form at the disc's inner rim and move inwards,
but aside from that, the surface density inside the inner cavity stays relatively low,
with values of about $100\,\text{M}_{\odot}\text{pc}^{-2}$, which is about 10 times 
lower than the surface density of the disc and about 4-6 times lower compared to the simulation
without magnetic field.

Thus the magnetic field seems to play an important role in maintaining a stable inner cavity
and in suppressing the mass flow into the inner cavity.
As the magnetic field is frozen into the plasma in the framework of ideal MHD, 
any chunk of material that moves towards the black hole deforms the otherwise toroidal magnetic field lines.
This deformation results in a restoring force, the magnetic tension force
\begin{equation}
 (\vec{B} \cdot \nabla)\vec{B}/4\pi \, ,
\end{equation}
that counteracts the inwards movement of matter and thus stabilizes the disc against radial collapse.

According to \citet{1981_Shivamoggi} the temporal growth rate of the Kelvin-Helmholtz instability in the presence of a
constant magnetic field that is parallel to the interface of the two fluids goes as $\exp \left( -i k c t\right)$
with wave vector $k$ and
\begin{equation}
 c = \frac{u}{2} \pm \sqrt{v_{\mathrm{A}}^2 - \frac{u^2}{4}} \,,
\end{equation}
where $u$ is the velocity difference of the fluids across the interface and $v_{\mathrm{A}}$ is the Alfv\'{e}n speed.
Thus a magnetic field has a stabilizing effect on the Kelvin-Helmholtz instability,
unless $u$ is more than twice the Alfv\'{e}n speed.
This might suppress the mixing of the disc material with the outflow and thus impede the extraction
of angular momentum from the disc.
This effect should be present everywhere in the disc where a toroidal magnetic field is present.

These results contradict previous knowledge in this field: magnetic fields are known to
enable accretion processes due to the magnetorotational instability \citep[MRI,][]{1991_Balbus_Hawley},
whereas in our simulations magnetic fields impede the accretion of gas.
Our results depend on the assumed initial condition of the magnetic field.
While the CND's magnetic field has been measured to be predominantly toroidal, it must also have small
poloidal (and radial) components, which are sufficient to drive the MRI.
Furthermore the growth rate of this instability is independent of the magnetic field strength,
i.e., only small perturbations are necessary to trigger the MRI.
Turbulent viscosity resulting from the MRI can potentially enhance accretion and might
mitigate or even negate the results reportet in this paper, but on time-scales that
are longer than the simulation time of the calculations presented in this paper.
Thus future investigations have to explore the effects of the MRI on the CND
and also test whether the results of the paper will hold on longer time-scales.

\begin{figure*}
 \makebox[\textwidth][c]{\includegraphics[width=\figwidth]{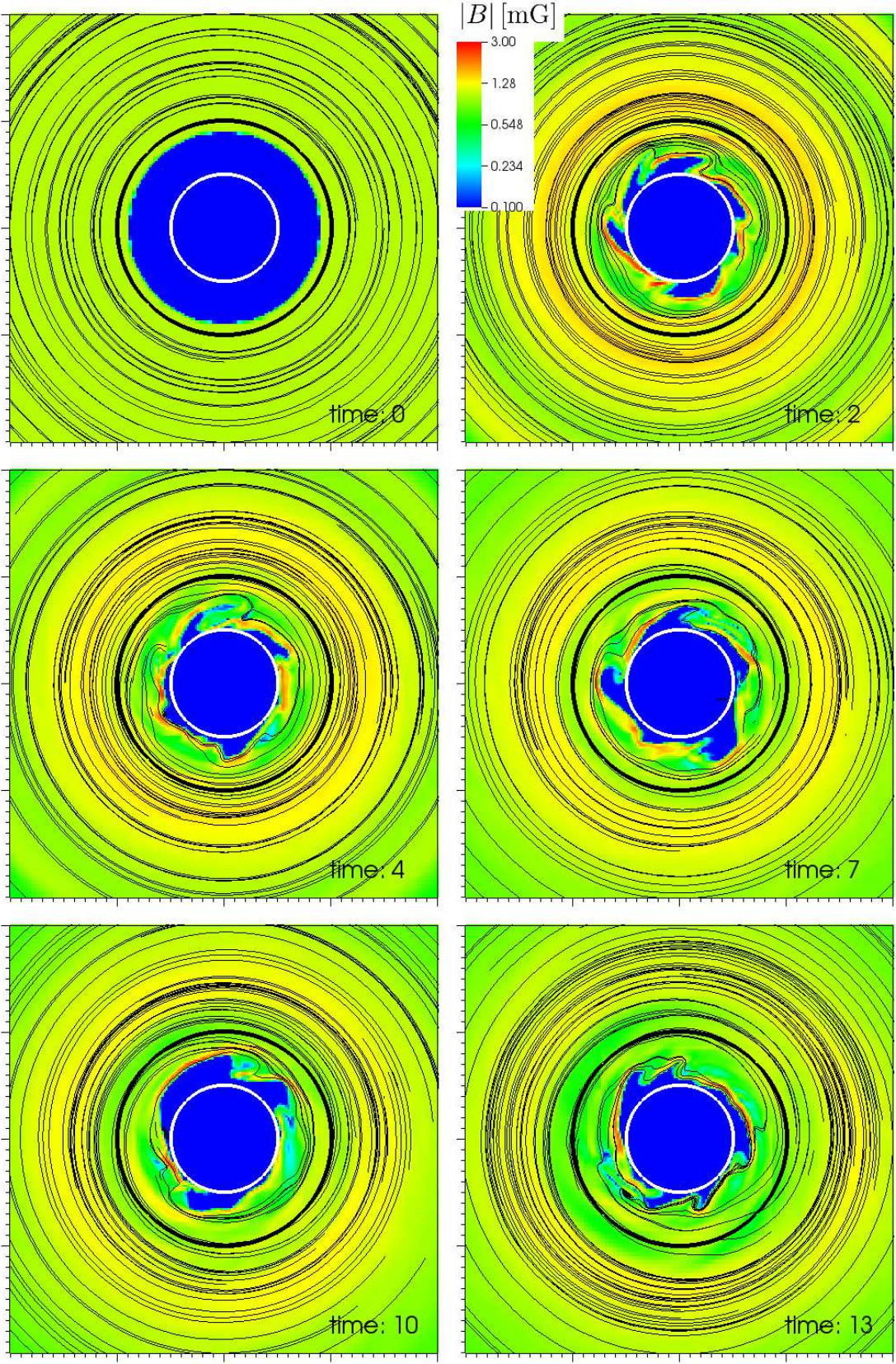}}
 \caption{Face-on view of the absolute value of the magnetic field strength at the disc's midplane
          for the simulation that includes magnetic fields for different times. Each snapshot shows a region of size $4{\times}4$\,pc.
          The time is given in units of the orbital time-scale at 1\,pc, which is $4.5\cdot10^4\,\text{yr}$.
          The thick black circle marks the location of the disc's initial inner rim,
          the thick white circle marks the region where the outflow is launched.
          Magnetic field lines are shown as thin black lines. Note that the color scale is logarithmic.}
 \label{fig:Bfield}
\end{figure*}

\begin{figure*}
 \makebox[\textwidth][c]{\includegraphics[width=\figwidth]{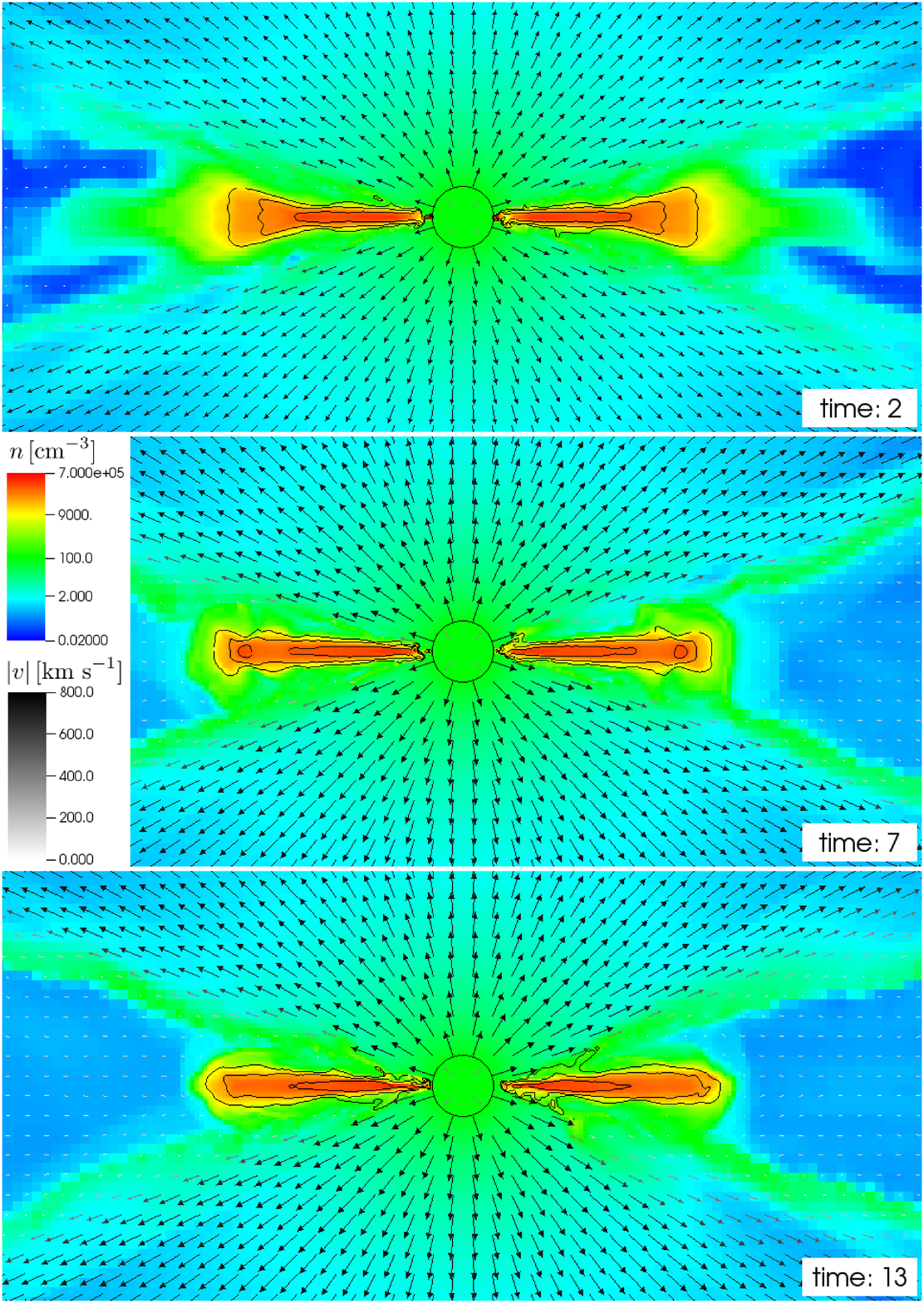}}
 \caption{Vertical cross section of the density for the simulation that includes magnetic fields.
          Each snapshot shows a region of size $15{\times}7$\,pc.
          The time is given in units of the orbital time-scale at 1\,pc, which is $4.5\cdot10^4\,\text{yr}$.
          The black circle marks the region where the outflow is launched.
          Contours of the magnetic field strength are shown as black lines,
          the arrows denote the velocity of the gas.}
 \label{fig:vertical}
\end{figure*}

In Fig.~\ref{fig:Bfield} we show the absolute value of the magnetic field with magnetic field lines.
The magnetic field basically keeps its initial toroidal configuration, and only inside the inner
cavity is it deformed by the clumps and streams of matter that are moving inwards. 
These streams have very high field strengths of 2-3\,mG,
because in the framework of ideal MHD the high density of the clumps implies a high magnetic energy density.
Fig.~\ref{fig:vertical} shows vertical cross sections for the simulation including magnetic fields.

\section{Angular momentum extraction}\label{sec:angmom}
The wind from the central stellar cluster interacts with the inner part of the CND.
To model the exchange of angular momentum,
we assume that this part is a ring of radius $s$ and radial extension $\Delta l$.
This is the radial depth into the disc that the wind penetrates and where an efficient
mixing of disc and wind material takes place.
We set $\Delta l = 0.1\,\mathrm{pc}$ as the instabilities shown in Fig.~\ref{fig:sigma_woB},
which provoke the mixing of disc and wind material, have approximately this scale.
Thus the ring has a mass of
\begin{equation}
 M_{\mathrm{D}} = 2 \pi \Sigma s \Delta l
 \label{eq:mass_ring}
\end{equation}
with the disc's surface density $\Sigma$, which is $950 \, \mathrm{M}_{\odot}\,\mathrm{pc}^{-2}$ for our choice of
initial conditions.

The ring has a total angular momentum of $J=j M_{\mathrm{D}}$ with specific angular momentum $j$.
The wind of the central stellar cluster adds mass with a rate of $f \dot{M}_{\mathrm{w}}$ to the ring,
where $\dot{M}_{\mathrm{w}} = 5 \cdot 10^{-3} \, \mathrm{M}_{\odot} \, \mathrm{yr}^{-1}$ is the total
outflow rate of the wind.
$f$ is the fraction of the wind that interacts with the ring,
which is the ratio of the surface of the disc's inner rim at $s$ to the surface of a sphere with radius $s$:
\begin{equation}
 f = \frac{2 \pi s h}{4 \pi s^2} = \frac{h}{2 s}
 \label{eq:fraction}
\end{equation}
$h=0.2\,\mathrm{pc}$ is the full thickness of the disc.
However, the total angular momentum of the ring does not change, as the outflow does not carry any angular momentum.
Thus after a time $dt$ the ring has the mass \mbox{$M_{\mathrm{D}} + f \dot{M}_{\mathrm{w}} dt$}
and the specific angular momentum
\begin{equation}
 j' = \frac{J}{M_{\mathrm{D}} + f \dot{M}_{\mathrm{w}} dt} = \frac{j}{1 + \frac{f \dot{M}_{\mathrm{w}}}{M_{\mathrm{D}}} dt} \, .
 \label{eq:angmom}
\end{equation}
Due to the loss of specific angular momentum the ring moves inwards to the radius $s-ds$.
Because \mbox{$j=\sqrt{G M_{\mathrm{BH}} s}$} and \mbox{$j'=\sqrt{G M_{\mathrm{BH}} \left( s-ds\right)}$}
eq.~\ref{eq:angmom} yields
\begin{equation}
 dt = \frac{M_{\mathrm{D}}}{f \dot{M}_{\mathrm{w}}} \left( \sqrt{\frac{s}{s-ds}} - 1 \right) = \frac{M_{\mathrm{D}}}{f \dot{M}_{\mathrm{w}}} \frac{ds}{2 s}
 \label{eq:time}
\end{equation}
where we used a Taylor expansion in the last step.

Inserting eqs.~\ref{eq:mass_ring} and \ref{eq:fraction} and integrating this equation gives the time the
inner rim needs to move inwards:
\begin{equation}
 \tau = \int\limits_{s_\text{i}}^{s_0} dt = \int\limits_{s_\text{i}}^{s_0} \frac{M_{\text{D}}}{f \dot{M}_{\mathrm{w}}} \frac{ds}{2 s} = \frac{2 \pi \Sigma \Delta l}{h \dot{M}_{\mathrm{w}}} \int\limits_{s_\text{i}}^{s_0} s ds = \frac{\pi \Sigma \Delta l}{h \dot{M}_{\mathrm{w}}} \left( s_0^2 - s_\text{i}^2\right) \,,
 \label{eq:timescale_wind}
\end{equation}
where $s_0=1\,\text{pc}$ is the disc's initial inner rim and $s_\text{i}=0.5\,\text{pc}$ the location of the rim after inwards migration.
Eq.~\ref{eq:timescale_wind} gives a time of \mbox{$\tau = 2.2 \cdot 10^5 \, \mathrm{yr} \approx 4.9 \, \tau_{\mathrm{orb}}$},
which fits well with the value of $4\,\tau_{\mathrm{orb}}$ found in section~\ref{sec:interact}.

We now calculate the radial velocity of the gas due to the angular momentum extraction:
\begin{equation}
 v_{\mathrm{r}} = \frac{ds}{dt} = \frac{2 s f \dot{M}_{\mathrm{w}}}{M_{\mathrm{D}}} = \frac{h \dot{M}_{\mathrm{w}}}{2 \pi s \Sigma \Delta l}
\end{equation}
The mass inflow rate is thus
\begin{equation}
 \dot{M} = 2 \pi s \Sigma v_{\mathrm{r}} = \frac{h \dot{M}_{\mathrm{w}}}{\Delta l} = 10^{-2} \, \mathrm{M}_{\odot} \, \mathrm{yr}^{-1} \,,
 \label{eq:massflow}
\end{equation}
which is 20 times the mass deposition rate from the outflow.
We compare this value to the mass inflow rate caused by the $\alpha$-viscosity $\nu_{\alpha} = \alpha c_{\mathrm{s}} h$.
We again use the values $c_{\mathrm{s}} \approx 1300\,\mathrm{m}\,\mathrm{s}^{-1}$ and $h=0.2\,\text{pc}$.
The inflow rate due to $\alpha$-viscosity is thus
\begin{equation}
 \dot{M}_{\alpha} \sim 3 \pi \nu_{\alpha} \Sigma = 3 \pi \alpha c_{\mathrm{s}} h \Sigma = \alpha \cdot 2.4 \cdot 10^{-3} \, \mathrm{M}_{\odot} \, \mathrm{yr}^{-1} \,.
 \label{eq:massflow_alpha}
\end{equation}
Even with a rather large value of $\alpha=1$ this is 4 times lower than the mass inflow of eq.~\ref{eq:massflow}.
However, the mass inflow of eq.~\ref{eq:massflow} will change with time, e.g., due to changes of the disc thickness $h$
or the extension of the mixing region $\Delta l$.
Furthermore at some distance from the black hole the gravitational potential of the central stellar
cluster will become relevant.
As its density is $\sim s^{-2}$, the angular momentum will rise linearly with $s$
outside the black hole's sphere of influence (i.e., beyond $s \sim$ 1-1.5 pc).
This will increase the time $dt$ of eq.~\ref{eq:time} by a factor of 2 and thus decrease
the mass inflow of eq.~\ref{eq:massflow} by a factor of 2.
All in all we can conclude that, even with a rather large value of $\alpha=1$,
the angular momentum extraction due to the wind is about as efficient as that due to $\alpha$-viscosity.
But whereas $\alpha$-viscosity is effective throughout the entire disc,
our mechanism is only effective at the inner rim of the disc where it interacts with the central outflow.

\section{A more detailed analysis}

In this section we present ten simulations of the circumnuclear disc.
In addition to the simulations shown in section~\ref{sec:interact}
we show simulations without magnetic field and without outflow,
with magnetic fields and without outflow, and with a lower initial
magnetic field strength of $B=0.1\,\text{mG}$ and with outflow.
Five of these simulations have an initial inner cavity of $r_{\text{in}}=1\,\text{pc}$
and five have an initial inner cavity of $r_{\text{in}}=2\,\text{pc}$.
The five different configurations of these simulations are summarized
in Table~\ref{tab:models}. The $r_{\text{in}}=1\,\text{pc}$ simulations cover a
time of $6 \cdot 10^5\,\text{yr}$, the $r_{\text{in}}=2\,\text{pc}$ simulations $10^6\,\text{yr}$. 
We emphasize that in the simulations without outflow there is no inner boundary,
i.e., the matter that flows inwards accumulates at the centre of the inner cavity.

\begin{table}
  \caption{Configurations of the simulations presented in this section.}
  \label{tab:models}
  \begin{tabular}{l|l|l|l}
    \hline
      No.   &  $B$ [mG]  &  outflow   &  linestyle (Figs.~\ref{fig:sigma1D_cav1}-\ref{fig:masses_cav2})   \\ \hline
      1.    &  0         &  yes       &  blue*, dotted        \\ 
      2.    &  1         &  yes       &  red*, dashed         \\ 
      3.    &  0         &  no        &  turquoise*, solid    \\ 
      4.    &  0.1       &  yes       &  green*, dashdotted   \\ 
      5.    &  1         &  no        &  black, solid         \\ 
   \hline
  \end{tabular} \\
  * grey in the printed version 
\end{table}

\begin{figure*}
 \makebox[\textwidth][c]{\includegraphics[width=\figwidth]{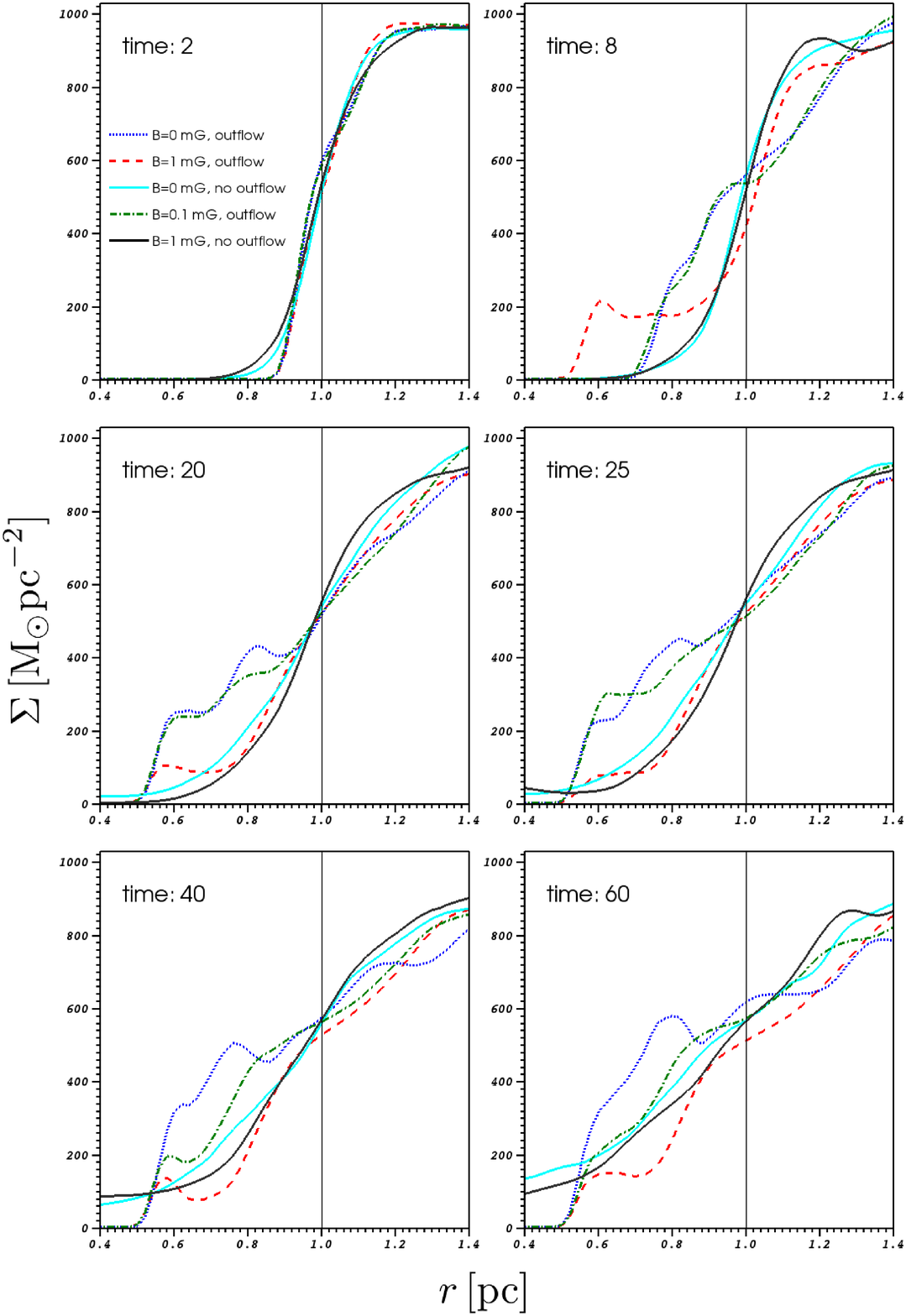}}
 \caption{Radial surface density profiles for the models presented in Table \ref{tab:models}.
          The vertical line marks the location of the disc's initial inner rim at 1\,pc.
          The time is given in units of $10^4\,\text{yr}$.
          The profiles have been calculated by using radial bins of the surface density with a bin size of 0.02\,pc.}
 \label{fig:sigma1D_cav1}
\end{figure*}

\begin{figure*}
 \makebox[\textwidth][c]{\includegraphics[width=\figwidth]{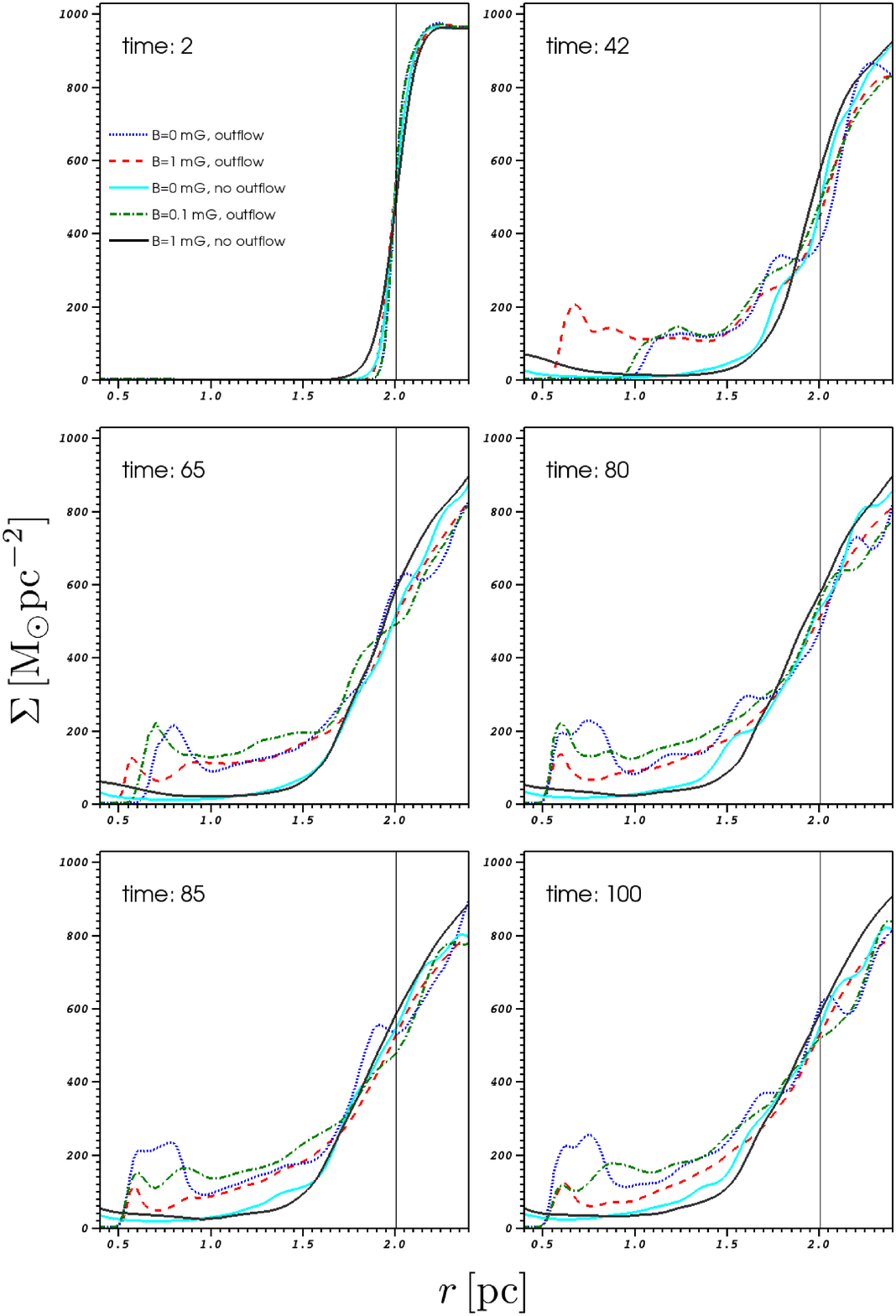}}
 \caption{Radial surface density profiles for the models presented in Table \ref{tab:models}.
          The vertical line marks the location of the disc's initial inner rim at 2 pc.
          The time is given in units of $10^4\,\text{yr}$.
          The profiles have been calculated by using radial bins of the surface density with a bin size of 0.02 pc.}
 \label{fig:sigma1D_cav2}
\end{figure*}

\begin{figure*}
 \makebox[\textwidth][c]{\includegraphics[width=\figwidth]{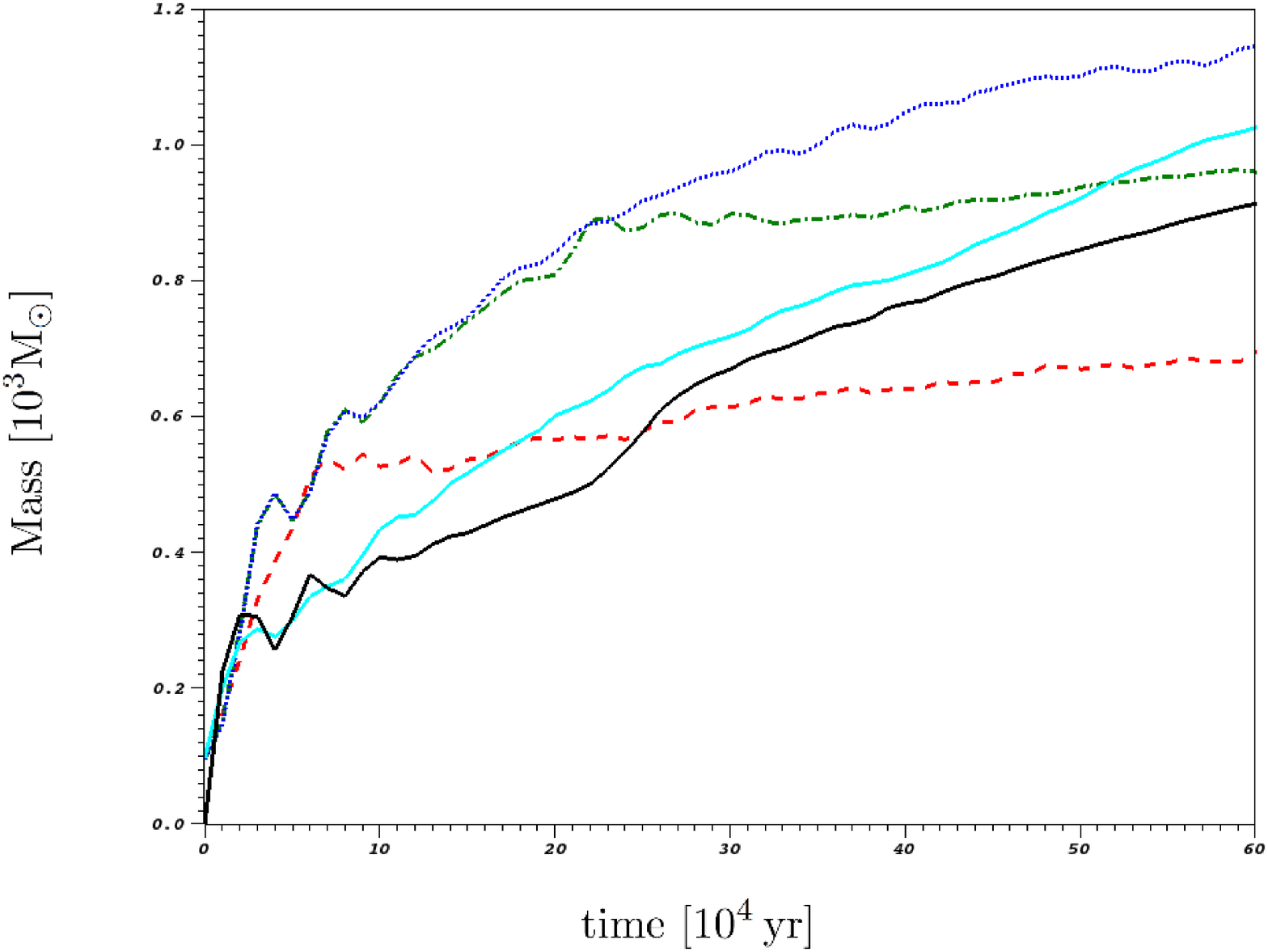}}
 \caption{Total mass within the inner cavity for an initial inner cavity of $r_{\text{in}}=1\,\text{pc}$
          and for the models presented in Table \ref{tab:models}.}
 \label{fig:masses_cav1}
\end{figure*}

\begin{figure*}
 \makebox[\textwidth][c]{\includegraphics[width=\figwidth]{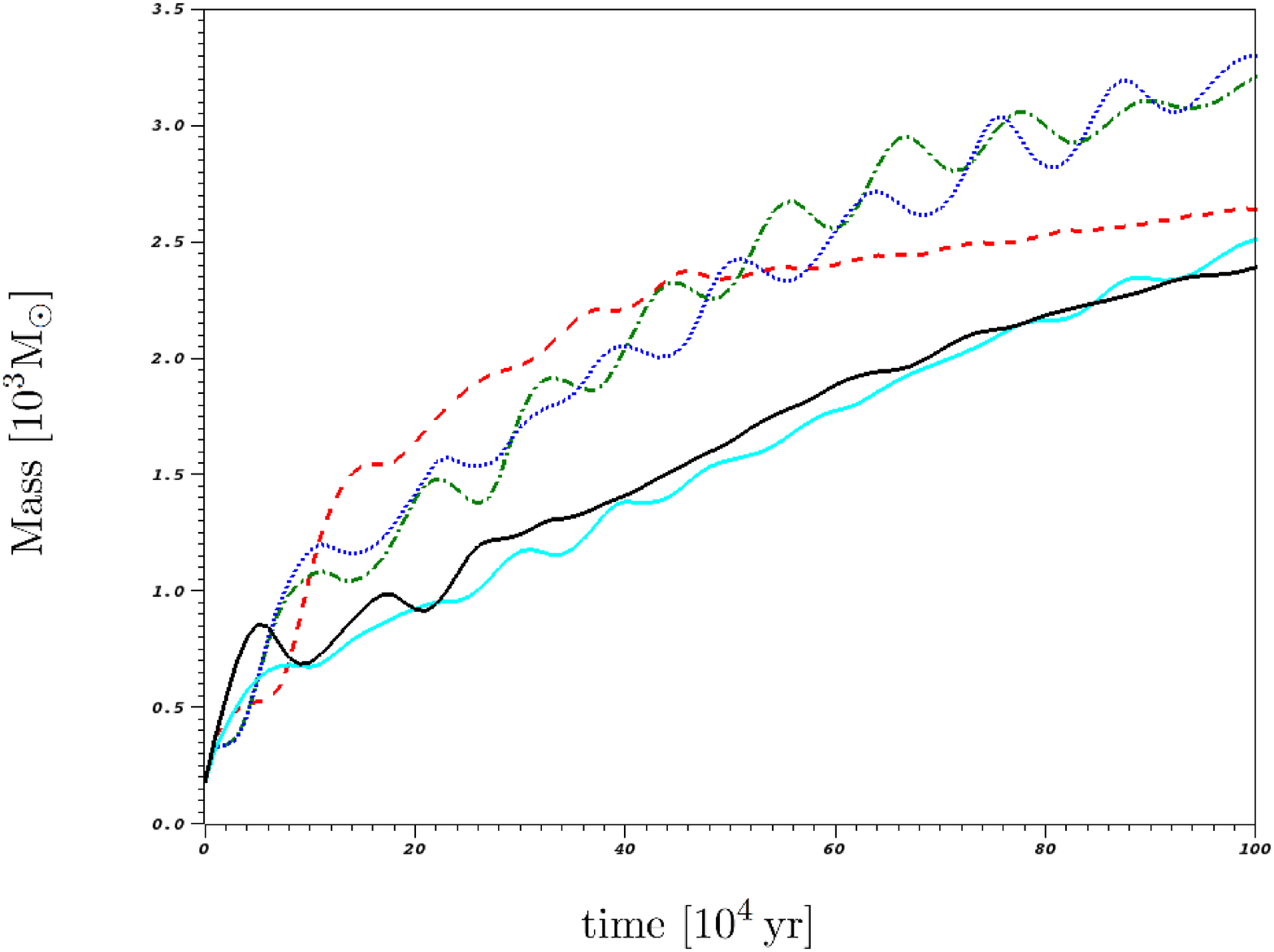}}
 \caption{Total mass within the inner cavity for an initial inner cavity of $r_{\text{in}}=2\,\text{pc}$
          and for the models presented in Table \ref{tab:models}.}
 \label{fig:masses_cav2}
\end{figure*}

Figs.~\ref{fig:sigma1D_cav1} and \ref{fig:sigma1D_cav2} show the disc's radial surface
density profiles of the $r_{\text{in}}=1\,\text{pc}$ and the $r_{\text{in}}=2\,\text{pc}$ results, respectively.
Figs.~\ref{fig:masses_cav1} and \ref{fig:masses_cav2} show the total mass within the inner cavity versus time
for an initial inner cavity of $r_{\text{in}}=1\,\text{pc}$ and $r_{\text{in}}=2\,\text{pc}$, respectively.

In all simulations the inner rim moves inwards in one way or another.
Different physical processes can basically be responsible for that, but they all work on different time-scales:
First, the inner rim moves inwards due to numerical viscosity on a viscous time-scale
\begin{equation}
 \tau_{\text{visc}} = \frac{2 \, r}{3 \, \nu_{\text{num}}} \, \Delta r
\end{equation}
where $\Delta r$ is the distance from the initial inner rim to the outflow object.
Second, gas pressure can cause the inner rim to move inwards on a time-scale of
\begin{equation}
 \tau_{P} = c_{\text{s}}^{-1} \, \Delta r \,,
\end{equation}
where $c_{\text{s}}$ is the speed of sound.
Furthermore, magnetic pressure can also cause the inner rim to migrate inwards on a time-scale
\begin{equation}
 \tau_{B} = v_{\text{A}}^{-1} \, \Delta r \,,
\end{equation}
where $v_{\text{A}}$ is the Alfv\'{e}n speed.
Finally, the inner rim moves inwards due to angular momentum extraction caused by its interaction with the wind,
on a corresponding time-scale given by eq.~\ref{eq:timescale_wind}.
All these time-scales are listed in Table~\ref{tab:timescales} for the
$r_{\text{in}}=1\,\text{pc}$ and the $r_{\text{in}}=2\,\text{pc}$ calculations, respectively.
$\tau_{\text{wind,ana}}$ is the time-scale for the angular momentum extraction according to eq.~\ref{eq:timescale_wind},
and $\tau_{\text{wind,sim}}$ is the time the inner rim actually needs to move inwards as found in the simulations.

\begin{table}
  \caption{Time-scales of relevant physical processes in the CND in units of $10^4$ yr.}
  \label{tab:timescales}
  \begin{tabular}{l|l|l}
    \hline
     time-scale                                   &   $r_{\text{in}}=1\,\text{pc}$   &   $r_{\text{in}}=2\,\text{pc}$   \\ \hline
     $\tau_{\text{orb}}$                         &   4.5    &     12.7    \\
     $\tau_{\text{visc}}$                        &   3190   &     19140   \\
     $\tau_{P}$                                  &   37.8   &     113.5   \\
     $\tau_{B}$ ($B=0.1\,\text{mG}$)             &   100    &     300     \\
     $\tau_{B}$ ($B=1\,\text{mG}$)               &   10     &     30      \\
     $\tau_{\text{wind,ana}}$                    &   22.5   &     112.5   \\
     $\tau_{\text{wind,sim}}$ ($B=0$)            &   20     &     65      \\
     $\tau_{\text{wind,sim}}$ ($B=1\,\text{mG}$) &   8      &     42      \\
   \hline
  \end{tabular}
\end{table}

At first we compare the simulations without outflow (No.~3 and 5).
The surface densities are similar for both configurations, and also the masses
within the inner cavity do not differ much; in the $r_{\text{in}}=1\,\text{pc}$ case there is slightly more
mass inside the inner cavity when magnetic fields are switched off.
Thus, in the absence of the central outflow, magnetic fields do not have a
large effect on the evolution of the CND, probably because initially there
is no magnetic field present inside the inner cavity.
Only numerical viscosity causes an inwards migration of matter.
For the $r_{\text{in}}=1\,\text{pc}$ simulations this process dominates the
mass balance within the inner cavity after about $20 \cdot 10^4\,\text{yr}$, but
for the $r_{\text{in}}=2\,\text{pc}$ simulations the mass balance inside the initial inner cavity
is not dominated by numerical viscosity during the time domain of our simulations.

Examining the $r_{\text{in}}=1\,\text{pc}$ simulations without magnetic fields (No.~1) and with a low initial magnetic
field strength (No.~4) shows that initially they do not deviate much from each other.
In both simulations the inner rim takes about $20 \cdot 10^4\,\text{yr}$ to reach the outflow object.
During the first $25 \cdot 10^4\,\text{yr}$ both simulations look more or less the same,
but after this time, both the surface density and the mass within the inner cavity are larger when
no magnetic fields are present.
Thus magnetic fields seem to suppress the inflow of matter into the inner cavity.
For the $r_{\text{in}}=2\,\text{pc}$ simulations without magnetic fields (No.~1) and with a low initial magnetic
field strength (No.~4) the inner rim takes about $65 \cdot 10^4\,\text{yr}$ to reach the outflow object.
For the first $80 \cdot 10^4\,\text{yr}$ the evolution of both simulations is similar,
and after this time the surface density is larger when no magnetic fields are present,
at least very close to the outflow object.
However, the total gas mass within the inner cavity does not show large differences between these two models.

In the simulation with magnetic fields and with outflow (No.~2) the inner rim moves inwards
very quickly, i.e., much faster than in the simulations without magnetic fields or with a low magnetic field strength.
It reaches the outflow object after about $8 \cdot 10^4\,\text{yr}$ ($42 \cdot 10^4\,\text{yr}$ for the $r_{\text{in}}=2\,\text{pc}$ simulation).
But after this time very little mass flows inwards, and the total mass inside the inner cavity and the
surface density are always smaller than compared to models No.~1 and 4.

From the simulations presented in this section we can draw two conclusions:
first, the outflow is causing the inner rim to move inwards, and the higher the magnetic field strength, the faster it moves.
One could assume that magnetic pressure drives the material inside,
because in the simulations with magnetic fields and with outflow (No.~2) the time-scales $\tau_{B}$ ($B=1\,\text{mG}$)
correspond to the times $\tau_{\text{wind,sim}}$ ($B=1\,\text{mG}$) the inner rim needs to move inwards.
But even in the simulations without magnetic fields (No.~1) the inner rim moves inward.
In that case, one could assume that this is caused by the gas pressure,
because $\tau_{P}$ also roughly corresponds to the times $\tau_{\text{wind,sim}}$ ($B=0\,\text{mG}$)
the inner rim needs to move inward.
However, the simulations without outflow do not show such an inward movement of the inner rim whatsoever.
Thus it is the outflow that causes the inner rim to shrink by being the main contributor to the
angular momentum extraction and the subsequent inflow of matter into the inner cavity.
The question of why the collapse is faster for higher magnetic field strengths
will be investigated in a forthcoming publication.

The second conclusion resulting from our investigation is that after the inner rim has reached
the outflow object, the surface density is lower for higher magnetic field strengths.
As we argued in section \ref{sec:interact}, this can be attributed to suppression of the inflow
into the inner cavity by the magnetic restoring force.
From the total mass within the inner cavity at the end of the simulations we can calculate
the average rate of mass flow into the cavity. The simulations without magnetic fields have mass inflow rates
of $1.9 \cdot 10^{-3} \, \mathrm{M}_{\odot} \, \mathrm{yr}^{-1}$ and $3.35 \cdot 10^{-3} \, \mathrm{M}_{\odot} \, \mathrm{yr}^{-1}$
for the $r_{\text{in}}=1\,\text{pc}$ and the $r_{\text{in}}=2\,\text{pc}$ simulations, respectively.
The simulations with magnetic fields ($B=1\,\text{mG}$) have mass flows of
$1.17 \cdot 10^{-3} \, \mathrm{M}_{\odot} \, \mathrm{yr}^{-1}$
and $2.65 \cdot 10^{-3} \, \mathrm{M}_{\odot} \, \mathrm{yr}^{-1}$
for the $r_{\text{in}}=1\,\text{pc}$ and the $r_{\text{in}}=2\,\text{pc}$ simulations, respectively.
These values are somewhat smaller than we would infer from eq.~\ref{eq:massflow},
but to get a comparable mass flow due to $\alpha$-viscosity we need a rather large value of $\alpha$,
of order of unity.

However, a part of this mass flow must be due to the angular momentum extraction of the wind.
Inflow of matter is also caused by numerical viscosity, as is evident from models No.~3 and 5
in Figs.~\ref{fig:masses_cav1} and \ref{fig:masses_cav2}.

\section{Clumping in the circumnuclear disc}\label{sec:clump}
As already shown in Section \ref{sec:interact} the interaction of the disc's inner rim with
the central outflow creates instabilities that lead to the formation of streams and clumps,
which then move inwards as they are sheared into the uniform medium surrounding them.

\begin{figure*}
 \makebox[\textwidth][c]{\includegraphics[width=\figwidth]{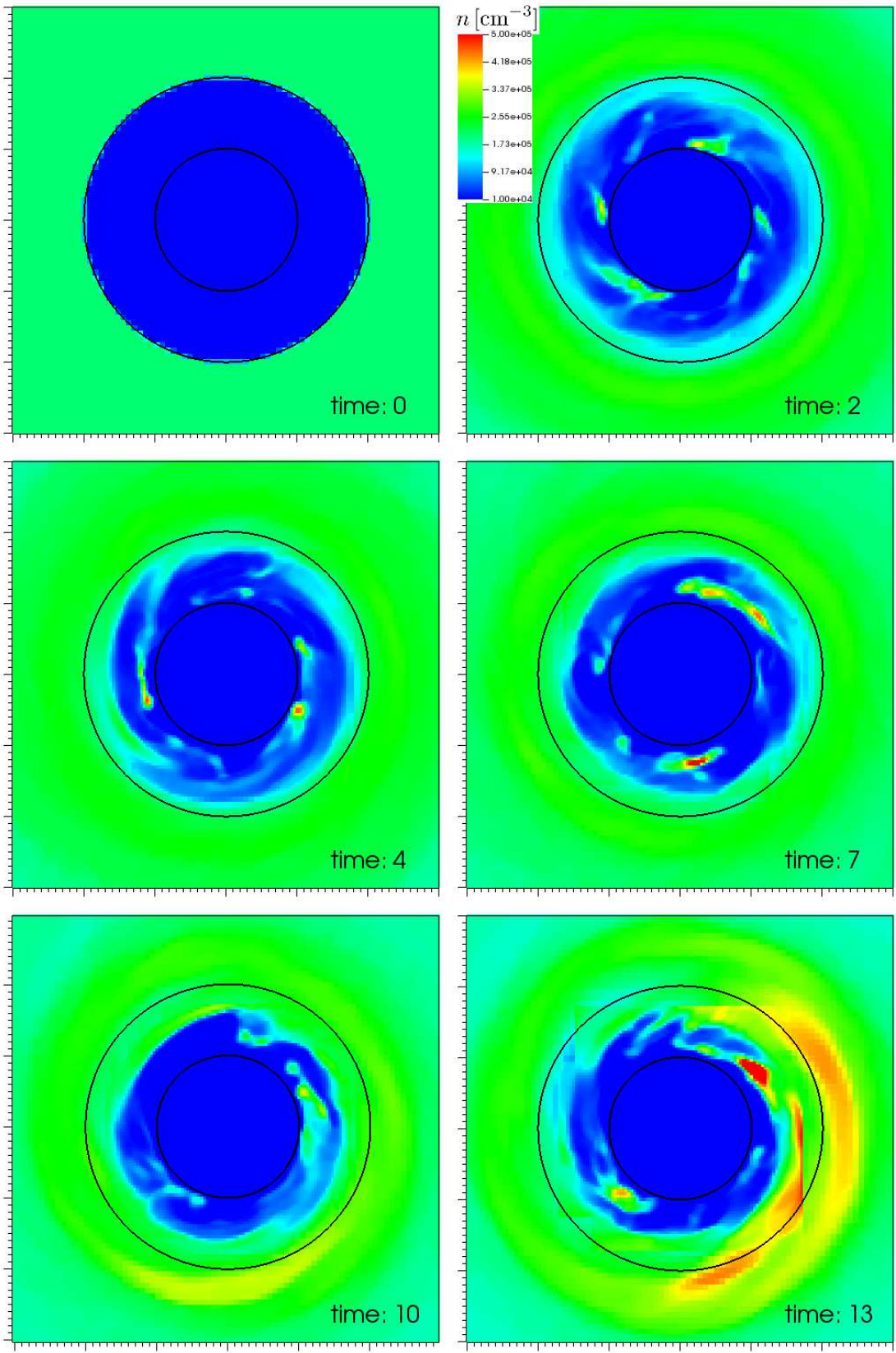}}
 \caption{Density at the disc's midplane in model No.~2 for different times.
          Each snapshot shows a region of size $3{\times}3$\,pc.
          The time is given in units of the orbital time-scale at 1\,pc, which is $4.5\cdot10^4\,\text{yr}$.
          The outer black circle marks the location of the disc's initial inner rim,
          the inner black circle marks the region where the outflow is launched.}
 \label{fig:dens}
\end{figure*}

\begin{figure*}
 \makebox[\textwidth][c]{\includegraphics[width=121mm]{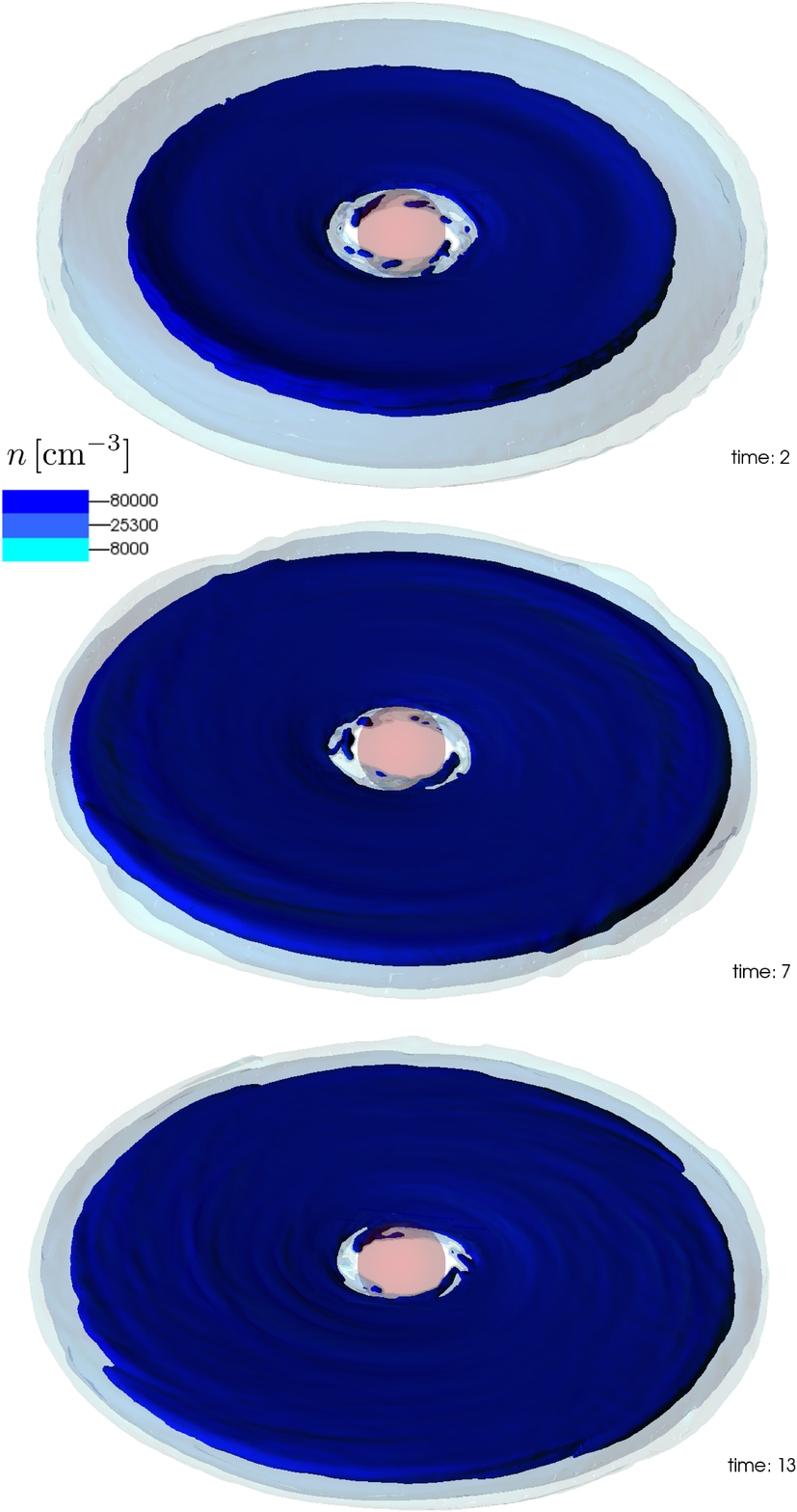}}
 \caption{Surfaces of constant density for the simulation that includes magnetic fields.
          The time is given in units of the orbital time-scale at 1\,pc, which is $4.5\cdot10^4\,\text{yr}$.
          The inner sphere marks the region where the outflow is launched.}
 \label{fig:contours}
\end{figure*}

In Fig.~\ref{fig:dens} we show the densities at the midplane of the disc for our model No.~2
(with $B=1 \,\text{mG}$ and outflow) at different times.
The clumps have densities of about $10^{5}\,\text{cm}^{-3}$ and sizes of $0.1-0.2\,\text{pc}$,
which correspond to the observed properties of the clumps in the CND.
However, these clumps are not stable against tidal shearing and thus get destroyed after a short period of time.
Fig.~\ref{fig:contours} furthermore shows surfaces of constant density for our model No.~2
and illustrates the clumpy structure of the disc's inner region.

Previous investigators have argued that the presence of unstable clumps in the vicinity of the black hole
implies that the CND is a transient feature.
Our results show that this need not necessarily be true, as we have demonstrated that such clumps
can be continuously created at the disc's inner rim due to an instability associated with its interaction with 
the central outflow.

Fig.~\ref{fig:dens} also shows the development of a spiral feature between 10 and 13 orbital time-scales.
This phenomenon will be discussed further in a forthcoming publication.

\section{Summary}\label{sec:concl}
We have investigated the interaction of the CND's inner rim with the outflow from the central stellar cluster.
The results of our simulations show that it is not the outflow that maintains the CND's inner cavity,
as had previously been assumed.
On the contrary, the interaction of the inner rim with the outflow causes the inner rim
to shrink within a few orbital time-scales.
However, including magnetic fields in our simulations suppresses the mass flow into the inner cavity and thus
supports the maintenance of a stable inner cavity, at least for the $10^6\,\text{yr}$ time-scale
over which we have run our simulations.
Whether magnetic fields can maintain the cavity on longer time-scales remains unclear.

Furthermore the interaction of the inner rim with the outflow creates instabilities that lead to the formation of clumps.
These clumps have the same properties as the clumps observed in the CND.
However, they are not stable against tidal shearing.
Thus the occurrence of unstable clumps in the CND does not necessarily mean that the disc is a transient feature,
since the presence of those clumps represents a balance between their continuous creation and their
disintegration by shearing.

Our results raise some important issues that we will address in our future work:
with a sufficiently high mass outflow rate in the central wind, the disc-wind interaction
can constitute an important method for angular momentum extraction from accretion discs
that is much more efficient than viscous torques, at least at the disc's inner rim.
In our simulation without a magnetic field, the inner rim collapses
within four orbital time-scales, whereas the time-scale for viscous processes is about 100 to 1000 orbital time-scales.
Our analytical estimation presented in Section \ref{sec:angmom},
based solely on angular momentum reduction by interaction with a radially outflowing wind,
gives also an accretion rate that is about as
large as an accretion rate caused by $\alpha$-viscosity.
However, our mechanism is not effective throughout the entire disc,
but only at its inner rim where it interacts with the central outflow.
This mechanism could also be relevant in other systems, for example, in protoplanetary discs around young stars,
if magnetic fields do not already impede accretion in such systems.
For both, protostellar and galactic centre systems, that poses the question of what magnetic
field strength is actually required to stabilize a disc from collapsing radially.
Likewise the role of viscous torques in this scenario is not yet clarified,
as they might increase the magnetic field strength required to stabilize the disc from collapsing.

\section*{Acknowledgements}
MB thanks the Fulbright Commission for support in the form of a foreign exchange scholarship.

\bibliographystyle{mn2e}
\bibliography{library}

\bsp

\appendix

\section{Numerical viscosity}\label{sec:numvisc}

\begin{figure}
 \makebox[0.47\textwidth][c]{\includegraphics[width=73mm]{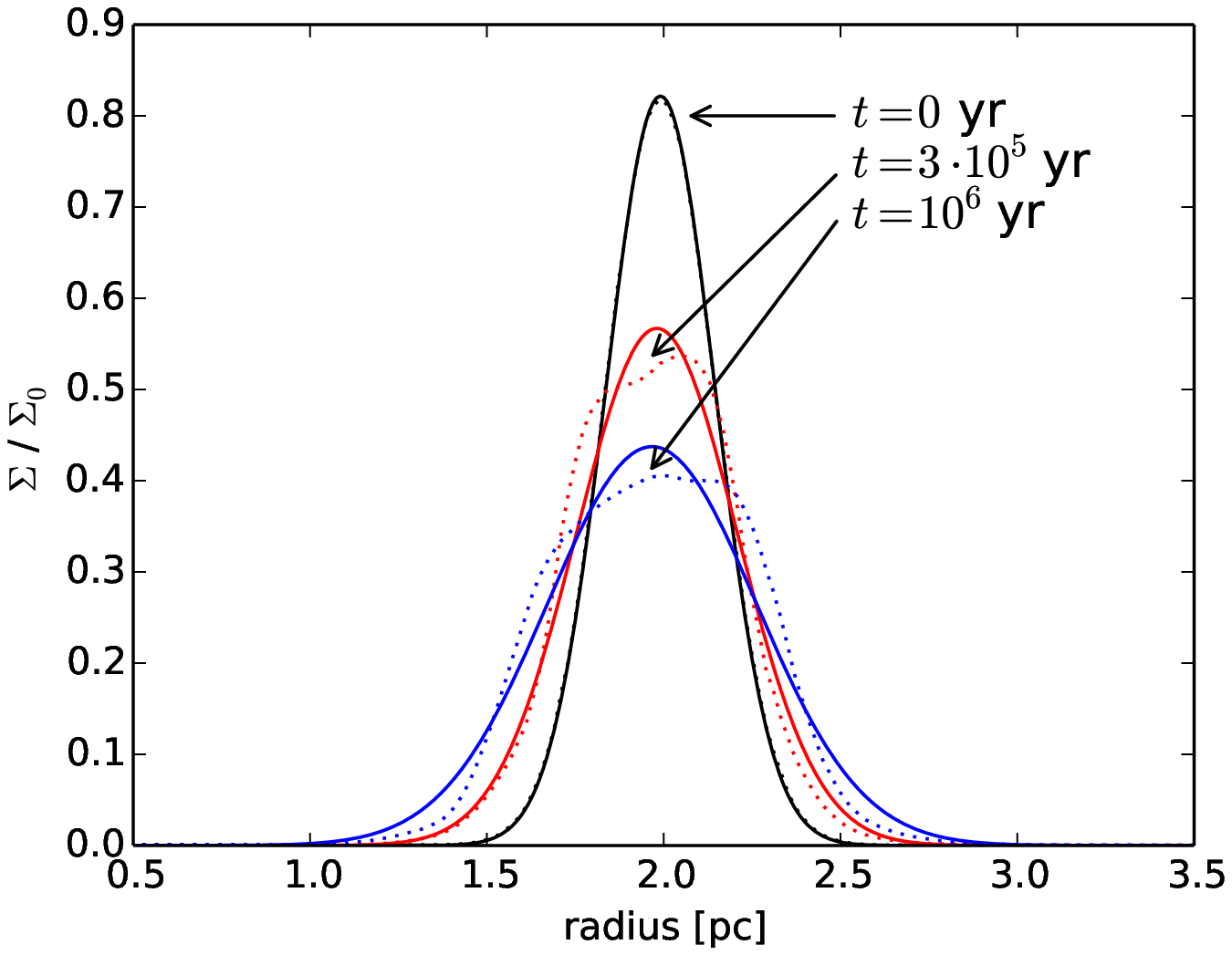}}
 \caption{Radial surface density profiles of the "spreading ring" problem for different times.
          Dotted lines: numerical solution, solid lines: fit to the numerical solution according to eq. \ref{eq:ring_analyt}.}
 \label{fig:spreadring_tau001}
\end{figure} 

In the following we determine the magnitude of numerical viscosity in disc simulations with AstroBEAR
and show that our simulations are not substantially affected by this effect.
Following \citet{1996_Murray} and \citet{2004_Lodato_Rice} we simulate the evolution of a
\textquotedblleft spreading ring\textquotedblright\, to measure the amount of numerical viscosity.

The \textquotedblleft spreading ring\textquotedblleft\, problem is based on the work of \citet{1974_Lynden-Bell_Pringle} and \citet{1981_Pringle}.
Assuming an accretion disc that is rotationally symmetric and geometrically thin and that
its angular frequency, $\omega$, does not change with time leads to the following equation that describes
the time-dependent evolution of an accretion disc:
\begin{equation}
 \frac{\partial \Sigma}{\partial t} + \frac{1}{r} \frac{\partial}{\partial r} \left[ \frac{\frac{\partial}{\partial r} \left( \nu \Sigma r^3 \frac{\partial \omega}{\partial r} \right)}{\frac{\partial}{\partial r} \left( r^2 \omega\right)} \right] = 0
 \label{eq:ring_evol}
\end{equation}
Here $\Sigma$ is the accretion disc's surface density and $\nu$ the viscosity of the gas.
For a Keplerian gravitational potential, constant viscosity and an
initial condition in the form of a delta peak at position $r_0$
\begin{equation}
 \Sigma \sim \delta (r-r_0) 
\end{equation}
eq.~\ref{eq:ring_evol} has the following analytical solution:
\begin{equation}
\Sigma (r,t) = \frac{\Sigma_0}{12 \pi r_0^2 x^{1/4} \tau} \text{exp} \left[ - \frac{(1+x^2)}{\tau} \right] I_{1/4} \left( \frac{2 x}{12 \tau} \right)
 \label{eq:ring_analyt}
\end{equation}
with
\begin{equation}
 x = \frac{r}{r_0} \, ,
\end{equation}
\begin{equation}
 \tau = \frac{\nu t}{r_0^2} \, ,
 \label{eq:tau}
\end{equation}
and $I_{1/4}$ is the modified Bessel function of the first kind.
As a delta peak is numerically difficult to handle we use eq.~\ref{eq:ring_analyt} as initial condition
with $r_0 = 2\,\text{pc}$ and $\tau=0.001$. 
The numerical and physical parameters are the same as in section~\ref{sec:ic} unless stated otherwise,
i.e., the resolution (cell size) at the location of the disc is $0.04\,\text{pc}$.
However, for our \textquotedblleft spreading ring\textquotedblright\, calculations we do not account for cooling,
magnetic fields and the outflow from the central stellar cluster.
The total simulation time is $10^6\,\text{yr}$.


Fig.~\ref{fig:spreadring_tau001} shows radial surface density profiles of the spreading ring for different times.
As expected the ring spreads out over time due to numerical viscosity.
To quantify the magnitude of numerical viscosity we fit a curve
according to eq.~\ref{eq:ring_analyt} to the radial surface density profile every $10^4$ yr.
This fit is shown as solid lines in Fig.~\ref{fig:spreadring_tau001}.
The only free parameter of the fit is $\tau$, thus the fitting procedure gives $\tau$ as a function of time.
Determining the slope of this function gives, according to eq.~\ref{eq:tau}, the viscosity $\nu$.

We define the parameter
\begin{equation}
\alpha_{\text{num}}=\frac{\nu}{\nu_{\alpha,\text{max}}} 
\end{equation}
where $\nu_{\alpha,\text{max}} = h c_{\text{s}}$ is the maximal $\alpha$-viscosity \citep{1973_Shakura_Sunyaev}.
Assuming that the CND has a temperature of $300\,\mathrm{K}$ and consists entirely of H$_{2}$, its speed of sound is
$c_{\mathrm{s}} \approx 1300\,\mathrm{m}\,\mathrm{s}^{-1}$.
For the disc thickness we use our initial condition $h=0.2\,\text{pc}$.
This results in a value of
\begin{equation}
  \alpha_{\text{num}}=(3.93 \pm 0.13) \cdot 10^{-2} \,.
\end{equation}
Considering that $\alpha$ is thought to have physical values of $\sim$~0.1-0.4 \citep[see, e.g.,][]{2007_King_Pringle_Livio},
and that the viscous time-scales derived in Table \ref{tab:timescales} are much larger than the total simulation
time and the time-scales of the inwards migration of the inner rim that we investigate in this paper,
this shows that our simulations are not dominated by numerical viscosity.


\begin{figure*}
 \makebox[\textwidth][c]{\includegraphics[width=147mm]{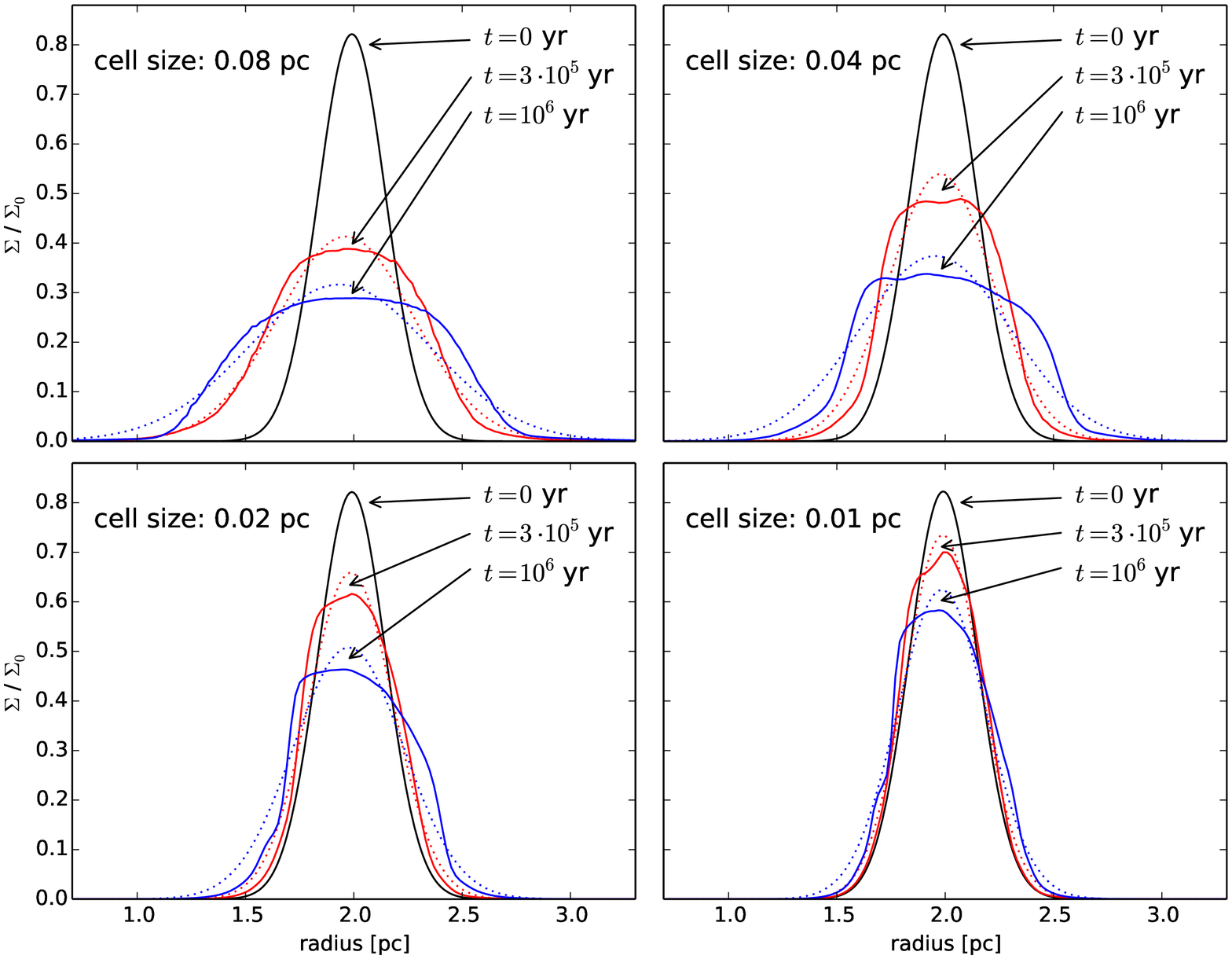}}
 \caption{Radial surface density profiles of the "spreading ring" problem for different times and different resolutions (cell sizes).
          Solid lines: numerical solution, dotted lines: fit to the numerical solution according to eq. \ref{eq:ring_analyt}.}
 \label{fig:spreadring_res}
\end{figure*}

We additionally conduct a resolution study with resolutions (cell sizes) of 0.08, 0.04, 0.02, and 0.01 pc.
However, we restrict these simulations to two dimensions to save computational time.
The evolution of the spreading ring for all resolution levels is shown in Fig. \ref{fig:spreadring_res}.
The qualitative evolution is similar for all the simulations, but as expected the spreading of the
ring is slower with higher resolution. The parameter $\alpha_{\text{num}}$ for all four simulations
is listed in Table \ref{tab:parameterstudy}.
The upper right figure with a resolution of 0.04 pc can be compared with the corresponding 3D simulation
(Fig. \ref{fig:spreadring_tau001}), again the qualitative evolution is similar, but the 2D simulation
has a higher numerical viscosity than the 3D simulation.
This resolution study furthermore shows that numerical viscosity is the main contributor of
the spreading of the ring and not any other process, like, e.g., pressure gradients.

\begin{table}
  \caption{Numerical viscosity for different resolution levels.}
  \label{tab:parameterstudy}
  \begin{tabular}{l|l}
    \hline
      resolution   &   $\alpha_{\text{num}} [10^{-2}]$    \\ \hline
      0.08 pc      &   $8.11 \pm 0.21$         \\
      0.04 pc      &   $6.22 \pm 0.05$         \\
      0.02 pc      &   $2.53 \pm 0.03$         \\
      0.01 pc      &   $1.14 \pm 0.02$         \\
   \hline
  \end{tabular}
\end{table}

\label{lastpage}

\end{document}